\documentclass{article}
\usepackage{amsmath,amssymb,amsfonts}
\usepackage{amsbsy}
\usepackage{graphicx}
\usepackage{cancel}
\usepackage{pgfplots}
\usepackage[subrefformat=parens,labelformat=parens]{subfig}
\usepackage{tikz}
\usepackage[compat=1.1.0]{tikz-feynman}

\newcommand{\red}[1]{{\color{red}#1}}

\def\m#1{\mathcal#1}

\newcommand{\be}{\begin{equation}}
\newcommand{\ee}{\end{equation}}

\def\bea{\begin{eqnarray}}
\def\eea{\end{eqnarray}}
\def\bean{\begin{eqnarray*}}
\def\eean{\end{eqnarray*}}
\def\cl{\centerline}

\def\head#1{$$gin{center}
\def\m#1{\mathcal#1}
\newcommand{\lambdab}{\boldsymbol\lambda}
\newcommand{\group}{$\mathcal{ PSL}_2(7)$ }
\newcommand{\more}{\vskip .2cm\hskip 4cm{\bf TO BE CONTINUED}\vskip .2cm}
\newcommand{\g}[1]{\mathbf{#1}}
\newcommand{\vev}[1]{\langle #1 \rangle}
\def\ov{\overline}
\shadowbox{
\begin{minipage}{3.0in}
\begin{center}
\bf \red{#1}
\end{center}
\end{minipage}}\end{center}\vspace{0.5\semcm} }
\begin{document}

\pagestyle{empty}
\cl{}

\cl{\LARGE The Freund-Rubin Coset, Textures  and Group Theory}
\vskip .5cm
\centerline {\Large Pierre Ramond}
\vskip .5cm 
\centerline {\large Institute for Fundamental Theory}
\centerline {\large University of Florida}
\vskip .5cm



\vskip 1cm
\cl {\Large In Memoriam  Peter G. O. Freund}
\vskip .5cm
\section{Introduction}

Forty years have elapsed since Freund and Rubin\cite{Freund} opened the way for the study of the seven-dimensional coset compact manifold that occurs in the descent of the eleven-dimensional supergravity to four dimensions. 

Can one find traces of this coset manifold  in the Standard Model? To any  group-inclined theorist, a compact seven-dimensional manifold is the domain of  the continuous group $G_2$, the smallest exceptional Lie group  well known as the automorphism group of Cayley numbers. 
 
Spurred by  eleven-dimensional physics, manifolds of $G_2$ holonomy  have been studied ever since the Freund-Rubin paper. These works concern themselves with continuous $G_2$. While (continuous) $SU_3$  is a subgroup of $G_2$, its discrete subgroups have not received the same attention.  

The search for an organizing pattern that explains the triplication of Standard Model's chiral families, their mixings and masses  initially suggested  the continuous family  symmetry $SU_3$. The  discovery of neutrino oscillations with  two large lepton mixing angles  pointed to  an underlying {\em Yukawa Crystallography} described by  a discrete family symmetry -- ergo a plethora of models\cite{Review,FlavorReviews} some with triplet representations of discrete subgroups of  $SU_3$.

 This author's research has concentrated on the discrete family symmetries generated by $\m T_7$, the  Frobenius group with 21 elements, and more recently{\cite{ramond1}  on $\m T_{13}$, the $39$-element  Frobenius group. The original physics motivation: both are $SU_3$ subgroups with triplet representations,  can unify quark and lepton flavor models.

The mathematical literature, offers an alternate {\em raison d'\^etre}  for $\m T_7$ as the maximal subgroup of  the simple group $ \m P\m S\m L_2(7)$ of order 168; 
 
Similarly, $\m T_{13}\times \m Z_2$ is the maximal subgroup of another simple group,  $\m P\m S\m L_2(13)$ of order 1092.
 
 Interestingly both  $\m P\m S\m L_2(7)$ and $\m P\m S\m L_2(13)$ are discrete subgroups of continuous $G_2$; both have seven-dimensional representations embedded in the septet of the  Lie algebra\cite{Slansky}  $G_2$.

 This paper    summarizes  well-known mathematical facts which may offer a discrete path from the Standard Model  to the Freund-Rubin $G_2$ manifold. 

\section{Some Frobenius Groups and Progenitors}
We begin with a description of the essentials of the Frobenius groups or order 21 and 39 and 78.
\vskip .5cm
 \noindent $\bullet$  $ \m T_7^{}=\m Z_7^{}\rtimes Z_3^{}$
 \vskip .5cm
\noindent  This 21-element Frobenius group   has two generators which satisfy the presentation 
 $$<S,\,T\,|\,S^3=T^7=e,~~S\,T\,S^2=T^2\,>$$
 It contains  five  conjugacy classes,
 
 $$
 C_1^{(1)}=\{\,e\,\},~ C_7^{(2)}=\{\,T\,\},~ C_7^{(3)}=\{\,S^2\,T\,S\,T^2\,\},
 \, C_7^{(4)}=\{\,T\,S^2\,\},~
  C_7^{(5)}=\{\,S\,\},
  $$
 with $(1,\,3,\,3,\,7,\,7)$ elements of order $(1,\,7,\,7,\,3,\,3)$,  respectively.

 Its five representations break into  three singlets representations:
 
 $${\bf 1}:~~~T=S=1,\quad {\bf 1'}:~~~T=1,\, S=\omega,\quad {\bf 1''}={\bf \bar 1'}:~~~T=1,\, S=\omega^2,\quad \omega^3=1,$$
 and one complex triplet representation with   conjugate:

$$S=\begin{pmatrix}0&1&0\\0&0&1\\ 1&0&0\end{pmatrix}\quad {\bf 3}:~T=\begin{pmatrix}\eta&0&0\\0&\eta^2&0\\0&0&\eta^4\end{pmatrix},\,~ 
{\bf \bar 3}:~T=\begin{pmatrix}\eta^6&0&0\\0&\eta^5&0\\0&0&\eta^3\end{pmatrix},
$$
where $\eta^7=1$.

 \vskip 1cm
  \noindent $\bullet$ $ \m T_{13}^{}=\m Z_{13}^{}\rtimes Z_3^{}$
 \vskip .5cm
 \noindent The 39-element Frobenius group  has  two generators with a similar presentation 
 
 $$<S,\,T\,|\,S^3=T^{13}=1,~~S\,T\,S^2=T^3\,>$$
 It contains seven conjugacy classes and  irreps ($n=0,1,\dots 12$):
 
 \bean
 C_1^{(1)}&=&\{\,e\,\},\quad C^{(2)}=\{\,S\,T^n\,\},\quad C^{(3)}=\{\,S^2\,T^n\,\}\quad C^{(4)}=\{\,T,\,T^3,\,T^9\,\},\\
  C^{(5)}&=&\{\,T^2,\,T^5,\,T^6\,\},\quad C^{(6)}=\{\,T^4,\,T^{10},\,T^{12}\,\},\quad C^{(7)}=\{\,T^7,\,T^8,\,T^{11}\,\}
 \eean
 with $(1,\,13,\,13,\,3,\,3,\,3,\,3)$ elements of order $(1,\,3,\,3,\,13,\,13,\,13,\,13)$ , respectively.
  
 \noindent  Its seven representations break into  three singlets representations:
 
 $${\bf 1}:~~~T=S=1,\quad {\bf 1'}:~~~T=1,\, S=\omega,\quad {\bf 1''}={\bf \bar 1'}:~~~T=1,\, S=\omega^2,\quad \omega^3=1,$$
 and two  inequivalent triplet representations  and their conjugates ($\rho^{13}=1$):
 
 \bean
 S=\begin{pmatrix}0&1&0\\0&0&1\\ 1&0&0\end{pmatrix}\quad {\bf 3}_1:~T&=&\begin{pmatrix}\rho&0&0\\0&\rho^3&0\\0&0&\rho^9\end{pmatrix},\,~ 
{\bf \bar 3}_1:~T=\begin{pmatrix}\rho^{12}&0&0\\0&\rho^{10}&0\\0&0&\rho^4\end{pmatrix},\\
{\bf 3}_2:~T&=&\begin{pmatrix}\rho^2&0&0\\0&\rho^6&0\\0&0&\rho^5\end{pmatrix},\, 
{\bf \bar 3}_2:~T=\begin{pmatrix}\rho^{11}&0&0\\0&\rho^7&0\\0&0&\rho^8\end{pmatrix}.
\eean

 \vskip .5cm 
 \noindent Their embeddings into continuous $SU_3$ are straightforward

 $$
 SU_3^{}\subset \m T^{}_7:\qquad {\bf 3}={\bf 3},\qquad {\bf 8}={\bf 1'}+{\bf \bar 1'}+{\bf 3}+{\bf \bar 3}
 $$
and two equivalent embeddings for $\m T_{13}$,
$$
SU_3^{}\supset \m T^{}_{13} :\qquad 
\begin{cases} {\bf 3}={\bf 3}_1,\quad {\bf 8}={\bf 1'}+{\bf \bar 1'}+{\bf 3_2}+{\bf \bar 3_2},\cr
  {\bf 3}={\bf 3}_2, \quad {\bf 8}={\bf 1'}+{\bf \bar 1'}+{\bf 3_1}+{\bf \bar 3_1}
 \end{cases}.$$

 \section{Frobenius groups and Spin Lattices}
 Textures based on Frobenius groups such as $\m T_7$ and $\m T_{13}$ can be viewed as  mappings between the world of data (Standard Model parameters) and a   potential $V(\varphi)$, depending on   local familons fields  $\varphi(x,\alpha)$, with only family charges (no gauged quantum numbers).  The familon potential is  invariant the family texture group.
  
They are labelled by  $x$  the space-time coordinate, and $\alpha$  the family symmetry index. From  the Freund-Rubin perspective, these two labels should unify into those of an eleven-dimensional manifold. 
 
Familons develop vacuum values which can be mapped backwards to the Standard Model and explain/predict its experimental consequences. This is the Familons's {\em ``Standard Model Portal"}. One may ask if there exist other  portals  into different physical systems, with the same  family symmetry expressed in terms of different physical variables, without the gauge {\em accoutrement}  of the Standard Model.
 \vskip .5cm
 
In 1981, it was noted\cite{Wyler}  that  nonAbelian discrete symmetries expressed as semi-direct products of cyclic Abelian groups, emerge  from spin lattice models with specialized couplings.  
\vskip .5cm
 \noindent Consider a square  lattice  with each lattice site labelled by ${\bf n}$. There  sits a ``spin" which can assume one of $p$ values where $p$ is prime  (for the groups discussed above $p=7, 13$), described by the Abelian $\m Z_p$ symmetry, a cyclic group with  $p$ one-dimensional representations with characters,
 
 $$\chi^{}_r=\exp\Big[\frac{2\pi i}{p}\,r\Big]\qquad r=0,1,...,p-1.$$
The state of the spin at each  lattice point ${\bf n}$ is labelled by the integers $\sigma_{\bf n}=\{0,1,2,...,p-1\}$.

Their interactions  between two lattice points ${\bf n}$ and ${\bf m}$, are determined by the Lagrangian,

$$L(\sigma_{\bf n},\sigma_{\bf m})=L(\sigma_{\bf n}-\sigma_{\bf m}),$$ 
which preserve global translation invariance over the whole  lattice. For simplicity interaction occurs only between nearest neighbors, with the Action  
 
 $$S=\sum_{<{\bf n}\,{\bf m}>} \, L(\sigma_{\bf n}-\sigma_{\bf m}).$$
Consider the transformations

$$\sigma_{\bf n}~\rightarrow ~\sigma_{\bf n}'=g(\sigma_{\bf n}),$$
 generated by the automorphism group of $\m Z_p$, which  maps the spins into themselves at each site. Global invariance  is achieved whenever 
 
 $$
 L(\sigma_{\bf n}-\sigma_{\bf m})~\rightarrow~L(g(\sigma_{\bf n})-g(\sigma_{\bf m})=L(\sigma_{\bf n}-\sigma_{\bf m}).
 $$
 When $p$  is an odd prime, the automorphism group of $\ Z_p$ is $\m Z_{p-1}$;   Its elements are powers of  generator(s) so that 
 
 $$g(\sigma_{\bf n})=v^m\sigma_{\bf n}\qquad \text{mod}(p)$$ 
 
\vskip .5cm 
 
 The other symmetry is the global translation (affine) symmetry under which  
 
 $$\sigma_{\bf n}~\rightarrow ~\sigma_{\bf n}+\tau\qquad \text{mod}(p)
 $$ 
 at all lattice sites, where $\tau$ is a fixed element of $\m Z_p$. 
 When combined, these two symmetries generate a nonAbelian discrete symmetry,   the semi-direct product of the two cyclic symmetries.

The generic  Action is then 

$$S(\{y_0,y_1,...,y_{p-1}\})=\sum_{<\bf n\,m>}\,\sum_{r=0}^{p-1}y_r^{}\exp\Big[\frac{2\pi i}{p}r(\sigma_{\bf n}-\sigma_{\bf m})\Big].
$$
The  global symmetry acts on the coupling constants  $y_0,y_1,...y_{p-1}$ which values determined by the desired symmetry.

\vskip .5cm

$\bullet$ Consider the case $p=7$ where the automorphism group is   $\m Z_6=\m Z_2\times \m Z_3$. Our construction yields three non-Abelian groups, ,  $\m Z_7\rtimes \m Z_3=\m T_7$, $\m Z_7\rtimes \m Z_2$, and $\m Z_7\rtimes \m Z_6$.
\vskip .3cm
For  $\m T_7$, let $v$ generate the homomorphic subgroup $\m Z_3$. The general action on the spins at each site is then

$$\sigma_{\bf n}~\rightarrow~v^m\sigma_{\bf n}+\tau, \qquad \text{mod}(7).$$ 
The integer $v$ is determined by the requirement 

$$v^3=1 \quad \text{mod}(7)~~\rightarrow~~\qquad \begin{cases} v=2: 2^3=8=1 \quad  \text{mod}(7)\\ v=4: 4^3=64=1 \quad  \text{mod}(7)\end{cases}
.$$
This symmetry is realized only for special values of the coupling constants, determined by the  $\m T_7$ representations, that is  $y_1=y_2=y_4$ and/or $y_3=y_5=y_6$.  
 \vskip .3cm
 The same construction with the full automorphism group yields $\m Z_7\rtimes \m Z_6=\m T_7\times \m Z_2$ which is realized only if all six coupling constants are equal.   
\vskip .5cm
 
 $\bullet$ If we take   $p=13$ the $\m X_{13}$ automorphism group is $\m Z_{12}=\m Z_6\times \m Z_2=\m Z_3\times \m Z_2\times \m Z_2$. The construction of the nonAbelian symmetries proceeds in the same way. The Action depends on thirteen coupling constants $y_0,y_1,...,y_{12}$.
 
 $\m T_{13}$ is constructed  by considering $\m Z_3$ as the automorphism group. Its generator $w$ is determined by requiring that
 
 $$w^3=1, \qquad \text{mod}(13)~~\rightarrow~~\qquad \begin{cases} w=3: 3^3=27=1 \quad  \text{mod}(13)\\ w=4: 9^3=729=1 \quad  \text{mod}(13)\end{cases}
.$$
Global invariance under $\m T_{13}$, is obtained whenever the twelve couplings  assemble themselves in four groups of triplets corresponding to $\bf 3_1$, $\bf 3_2$ and their conjugates. 
 \vskip .3cm
By choosing   $\m Z_6$ as  the automorphism group, the same construction realizes the  $\m T_{13}\times \m Z_2$ symmetry by grouping  the couplings into groups of six with equal values, reflecting its two sextet representations.

 \section{The Road to $G_2$}
 Searching  for a theoretical origin of these Frobenius symmetries is tantamount to finding   their progenitor simple groups\cite{ATLAS}. We begin with some mathematical factoids:
\vskip .5cm

- $\m T^{}_7$ is the largest  maximal subgroup of $ \m P\m S\m L_2(7)$: 
 \vskip .3cm
 $$\m P\m S\m L_2(7)\supset \m T^{}_7. $$
 \vskip .5cm

- $\m T^{}_{13}$ is naturally embedded into the $78$-element Frobenius group $\m Z_{13}^{}\rtimes Z_6^{}$, itself embedded into   $ \m P\m S\m L_2(13)$, the  simple group of order $1092$:
\vskip .3cm
$$ \m P\m S\m L_2(13)~\supset~\m Z_{13}^{}\rtimes Z^{}_6~\supset~\m T^{}_{13}$$

\vskip .5cm 
\noindent These two simple groups have a common feature:  real seven-dimensional representations, which suggests that they are embeddable into continuous $G_2$.
\vskip .5cm 
 \noindent This is indeed the case, as shown by various authors\cite{Wales,Meurman,Wybourne}

\noindent  A list of $G_2$'s  seven irreducible  discrete subgroup can be be found in\cite{EvansPugh,He}; as expected, all  have real septet representations.

To begin, note that  the $G_2$ Kronecker product of two fundamentals  

$$
{\bf 7}\times{\bf 7}=\big[{\bf 7}+{\bf 14}\big]_a+\big[{\bf 1}+{\bf 27}\big]_s
$$
 contains the septet in its antisymmetric product.   Therefore a necessary condition for a good embedding is that the Kronecker product of the subgroup representations must satisfy this requirement, thus limiting  the possible embeddings. The rest of the decomposition  in the antisymmetric product then expresses  the  $G_2$ adjoint in terms of the subgroup's representations.

We now summarize  from  Evans and Pughs\cite{EvansPugh}  the specific embeddings for each of the seven cases.

\vskip .75cm

$\bullet$   {\large $ G_2~\supset~ \m P\m S\m L_2(7)$}
\vskip .5cm 
\noindent Order 168. Irreps: ${\bf 1},{\bf 3},{\bf  \bar 3}, {\bf 6},{\bf 7},{\bf 8}$. 
\vskip .3cm 
\noindent Kronecker products:  

${\bf 7}\times {\bf 7}=({\bf 7}+{\bf 3}+{\bf \bar 3}+{\bf 8})_a+({\bf 1}+{\bf 6}+{\bf 6}+{\bf7}+{\bf 8})_s$

$
({\bf 1}+{\bf 3}+{\bf \bar 3})\times({\bf 1}+{\bf 3}+{\bf \bar 3})=({\bf 1}+{\bf 3}+{\bf \bar 3}+{\bf 3}+{\bf \bar 3}+{\bf 8})_a+
({\bf 1}+{\bf 6}+{\bf 6}+{\bf 7}+{\bf 8})_s$

\vskip .3cm 
\noindent Two  embedding: 

${\bf 7}={\bf 7},\qquad {\bf 14}={\bf 3}+{\bf \bar 3}+{\bf 8}$

${\bf 7}={\bf 1}+{\bf 3}+{\bf \bar 3},\qquad {\bf 14}={\bf 3}+{\bf \bar 3}+{\bf 8}$
\newpage
$\bullet$   {\large $G_2~\supset~ \m P\m G\m L_2(7)$}
\vskip .5cm
\noindent Order 336. Irreps: ${\bf 1},{\bf 1}_1,{\bf 6}_1,{\bf   6}_2, {\bf 6}_3,{\bf 7}_1,{\bf 7}_2,{\bf 8}_1,{\bf 8}_2$.
\vskip .3cm 
\noindent Kronecker product: 

${\bf 7_2}\times {\bf 7}_2=({\bf 7_2}+...)_a+({\bf 1}+\cdots )_s$

$({\bf 1}_1+{\bf 6}_1)\times ({\bf 1}_1+{\bf 6}_1)=(({\bf 1}_1+{\bf 6}_1)+{\bf 1}'+{\bf 6}_1+{\bf 8}_1)_a+({\bf 1}+{\bf 6}_1+{\bf 6}_2+{\bf 6}_3+{\bf 8}_2)_s$
\vskip .3cm 
\noindent Two embeddings: 

${\bf 7}={\bf 7}_2, \qquad {\bf 14}={\bf 6}_1+{\bf 8}_1$

${\bf 7}={\bf 1}_1+{\bf 6}_1,\qquad {\bf 14}=....$
\vskip .75cm 

$\bullet$   {\large$G_2~\supset~ \m P\m S\m L_2(7)\rtimes(\m Z_2\times\m Z_2\times\m Z_2)$}
\vskip .5cm 

\noindent Order 1344. Irreps: ${\bf 1},{\bf 3},{\bf  \bar 3}, {\bf 6},{\bf 7}_1,{\bf 7}_2,{\bf 7}_3,{\bf 8},{\bf 14},{\bf 21}_1,{\bf 21}_2$. 
\vskip .3cm 
\noindent Kronecker product: ${\bf 7_1}\times{\bf 7}_1=({\bf 7_1}+{\bf 14})_a+({\bf 1}+{\bf 6}+{\bf 21})_s$
\vskip .3cm 
\noindent Two embeddings: ${\bf 7}={\bf 7}_1, \quad \text{ and} \quad {\bf 7}={\bf 7}_2 $


\vskip .75cm 

$\bullet$   {\large $G_2~\supset~ \m P\m S\m L_2(8)$}
\vskip .5cm 
\noindent Order 504. Irreps: ${\bf 1},{\bf  7}_1, {\bf 7}_2,{\bf 7}_3,{\bf 7}_4,{\bf 8},{\bf 9}_1,{\bf 9}_2,{\bf 9}_3$. 
 \vskip .3cm 
\noindent Kronecker product: ${\bf 7_1}\times{\bf 7}_1={\bf 7_2}\times{\bf 7}_2=({\bf 7_1}+{\bf 7}_2+{\bf 7}_3)_a+({\bf 1}+{\bf 9}_1+{\bf 9}_2+{\bf 9}_3)_s$

\vskip .3cm 
\noindent Two embeddings: ${\bf 7}={\bf 7}_1, \quad {\bf 14}={\bf 14}_1 $  or  ${\bf 7}={\bf 7}_2, \quad {\bf 14}={\bf 14}_1  $

 \vskip .75cm 

$\bullet$   {\large $G_2~\supset~ \m P\m S\m L_2(13)$}
\vskip .3cm 
 \noindent Order 1092. Irreps: ${\bf 1},{\bf  7}_1, {\bf 7}_2,{\bf 12}_1,{\bf 12}_2,{\bf 12}_3,{\bf 13},{\bf 14}_1,{\bf 14}_2$.
  \vskip .5cm 
\noindent Kronecker product: 

${\bf 7_1}\times{\bf 7}_1=({\bf 7_1}+{\bf 14}_1)_a+({\bf 1}+{\bf 13}+{\bf 14_2})_s$

${\bf 7_2}\times{\bf 7}_2=({\bf 7_2}+{\bf 14}_1)_a+({\bf 1}+{\bf 13}+{\bf 14_2})_s$
\vskip .3cm 
\noindent Two embeddings: ${\bf 7}={\bf 7}_1, \qquad {\bf 14}={\bf 14}_1$   or ${\bf 7}={\bf 7}_2, \qquad {\bf 14}={\bf 14}_1$
  \vskip .75cm 
$\bullet$   {\large $G_2~\supset~ PU(3;3)$}
\vskip .5cm 
\noindent Order: 6048. Irreps: ${\bf 1},{\bf 6},{\bf  7}_1, {\bf 7},{\bf \bar 7},{\bf 14},{\bf 21}_1,{\bf 21},{\bf \overline {21}},{\bf 27},{\bf 28},{\bf \overline {28}},{\bf 32},{\bf \overline { 32}}$.
 \vskip .3cm 
\noindent Kronecker product: ${\bf 7}\times{\bf 7}=({\bf 7}+{\bf 14})_a+({\bf 1}+{\bf 27})_s$
\vskip .1cm 
\noindent  One  embedding: ${\bf 7}={\bf 7}$

\vskip 1cm 

$\bullet$   {\large$G_2~\supset~ G_2(\m {Z}_2)$}
\vskip .3cm
  \noindent Order: 12096. Irreps:  ${\bf 1},{\bf 1}_1,{\bf 1}_2,{\bf 6},{\bf \bar 6},{\bf 7}_1,{\bf 7}_2,{\bf  14}_1, {\bf 14}_2,{\bf 14}_3,{\bf 21}_1,{\bf 21}_2,{\bf 27}_1,{\bf 27}_2,{\bf 42},{\bf 56},{\bf 64}$.

 \vskip .5cm 
\noindent Kronecker product: ${\bf 7}\times{\bf 7}=({\bf 7}+{\bf 14})_a+({\bf 1}+{\bf 27})_s$

\vskip .3cm 
\noindent  One embedding: ${\bf 7}={\bf 7},\quad {\bf 14}={\bf 14}$
\vskip .5cm

\noindent The interested reader can find more information in\cite{EvansPugh}, especially on the McKay graphs which provide a graphical rendion of the Kronecker products.

\vskip .3cm

\noindent We mention for completeness the embeddings of $\m Z_{13}^{}\rtimes Z^{}_6$,  with two real sextuplet representations into  $ \m P\m S\m L_2(13)$, 
\vskip .2cm
${\bf 7}_1={\bf 1}+{\bf 6_1}={\bf 1}+{\bf 3}_1+{\bf \bar 3}_1;\qquad {\bf 7}_2={\bf 1}+{\bf 6_2}={\bf 1}+{\bf 3}_2+{\bf \bar 3}_2$
\vskip .3cm
It is not clear if any of these mathematical connections will prove relevant to an understanding of the roots of the Standard Model. It is nevertheless helpful to gather some of the information in on place which perhaps will bear fruit.


\section{Englert Fluxes and $ \m P\m S\m L_2(7)$}
I have to  mention the work of Pietro Fr\'e and collaborators\cite{Fre}  who compactify eleven-dimention supergravity to
 a four-dimensional space-time with   M-2 branes. Their Freund-Rubin manifold is  generalized to a kind of squashed sphere of the type introduced by Englert\cite{Englert}, where the three form satisfies a complicated equations. 
 
 They construct solutions for the three-form on  a seven-dimensiona' manifold modded out by a ``crystallographic lattice" which they take to be $ \m P\m S\m L_2(7)$. They find a supersymmetric solution  with the Frobenius symmetry $\m T_7$.
 
 Their work provides  yet another example where these nonAbelian discrete symmetries arise.  In fact they believe that it can shed light on M-theory.
 
 It would be very interesting if their construction generalizes to $\m T_{13}$, or to any of the finite subgroups of $G_2$ described above.

 \section{A personal note}
 I met Peter in person soon after arriving at FermiLab (then NAL). My early recollection  was at a seminar by  Nambu-Sensei:   received in silence by an attentive audience desperately trying to understand the wisdom of the master, until a booming voice arose ``Yoichiro Yoichiro, no, no,no ...", and rushing to the blackboard to explain; I do not remember the rest. My friend  Lou Clavelli who had been a student at Chicago told me, this is Peter Freund.
 
This stentorian  voice belonged to an extremely friendly and inquisitive person, who knew an amazing amount of physics,  mathematics, and  not least stories! A font of knowledge on many fronts.
 
 The last time I saw Peter was at the Nambu memorial in Osaka, where we shared a few days which  I very much enjoyed in his civilized  company. 
 
 I am honored by writing this paper on one of his fundamental contributions to physics.
 \vskip 1cm
 
 \noindent I  thank my collaborators for useful discussion with my students,  Moinul Rahat and Bin Xu, and  acknowledge partial support from the U.S. Department of Energy under award number DE-SC0010296.


{}

\end{document}